\documentclass[runningheads]{llncs}
\usepackage{makecell}
\usepackage{graphicx}
\usepackage{amssymb}
\usepackage{framed,multirow,color}
\usepackage{marvosym}
\begin{document}
\title{Structure-consistent Restoration Network for Cataract Fundus Image Enhancement}
\titlerunning{SCR-Net for Cataract Fundus Image Enhancement}
%
\author{Heng Li\inst{1,}\textsuperscript{\Letter} \and
Haofeng Liu\inst{1} \and
Huazhu Fu\inst{2} \and
Hai Shu\inst{3} \and
Yitian Zhao\inst{4} \and
Xiaoling Luo\inst{5} \and
Yan Hu\inst{1,}\textsuperscript{\Letter} \and
Jiang Liu\inst{1,6,7}}
%
\institute{Department of Computer Science and Engineering, Southern University of Science and Technology, Shenzhen, China \\
\email{lih3, huy3@sustech.edu.cn}\and
IHPC, A*STAR, Singapore\and
Department of Biostatistics, School of Global Public Health, New York University, New York, USA\and
Cixi Institute of Biomedical Engineering, Chinese Academy of Sciences, China\and
Shenzhen People's Hospital, Shenzhen, China\and
Guangdong Provincial Key Laboratory of Brain-inspired Intelligent Computation, Southern University of Science and Technology, Shenzhen, China\and
Research Institute of Trustworthy Autonomous Systems, Southern University of Science and Technology, Shenzhen, China}
\maketitle              
\begin{abstract}
Fundus photography is a routine examination in clinics to diagnose and monitor ocular diseases. However, for cataract patients, the fundus image always suffers quality degradation caused by the clouding lens. The degradation prevents reliable diagnosis by ophthalmologists or computer-aided systems. To improve the certainty in clinical diagnosis, restoration algorithms have been proposed to enhance the quality of fundus images. 
Unfortunately, challenges remain in the deployment of these algorithms, such as collecting sufficient training data and preserving retinal structures. 
In this paper, to circumvent the strict deployment requirement, a structure-consistent restoration network (SCR-Net) for cataract fundus images is developed from synthesized data that shares an identical structure.
A cataract simulation model is firstly designed to collect synthesized cataract sets (SCS) formed by cataract fundus images sharing identical structures. 
Then high-frequency components (HFCs) are extracted from the SCS to constrain structure consistency such that the structure preservation in SCR-Net is enforced.
The experiments demonstrate the effectiveness of SCR-Net in the comparison with state-of-the-art methods and the follow-up clinical applications. The code is available at https://github.com/liamheng/ArcNet-Medical-Image-Enhancement.
 
\keywords{Cataract, fundus image enhancement, high-frequency components, structure consistency.}
\end{abstract}

\section{Introduction}
High-quality medical images are the foundation for modern diagnosis and monitoring.
The advantages in safety and cost have made fundus photography become a routine examination in clinics to diagnose and monitor ocular diseases~\cite{li2021applications,zhang2020machine}. 
More recently, automatic algorithms have been developed to assist the clinical screening and diagnosis based on fundus images~\cite{wang2019patch,zhang2019automated,zhao2019direct}. 
Unfortunately, as a pathological characteristic, low quality is unavoidable for the fundus images collected from cataract patients.
The degradation in cataract images not only impacts the performance of automatic algorithms, but also prevents reliable diagnosis by ophthalmologists~\cite{macgillivray2015suitability,ZHANG2022104037}.
Therefore, cataracts lead to uncertainty in clinical observation and the risk of misdiagnosis.

To improve the certainty in ophthalmic diagnosis and treatment, the medical imaging community has strived to overcome the degradation of fundus images caused by missing focus, uneven illumination, as well as cataracts~\cite{li2021applications}.
Histogram equalization~\cite{mitra2018enhancement}, spatial filtering~\cite{cheng2018structure}, and frequency filtering~\cite{cao2020retinal} were imported to develop fundus image enhancement algorithms. 
However, these methods are not sensitive to retinal details or not generalizable in clinical scenarios. 
In recent years, deep learning has been employed to adaptively learn restoration models for fundus images~\cite{chen2018robust,huang2021neighbor2neighbor,lore2017llnet}. 
To overcome the limitation caused by the requirement of vast supervised data, unsupervised algorithms and data augmentation techniques have been developed to implement fundus image restoration.
CycleGAN~\cite{zhu2017unpaired} and contrastive unpaired translation (CUT)~\cite{park2020contrastive} were modified ~\cite{cheng2021secret} to learn suitable mappings from a low-quality domain to a high-quality domain from unpaired data. 
Alternatively, low-high quality paired data were synthesized for the training of fundus enhancement networks~\cite{liu2022domain,luo2020dehaze,shen2020modeling}. 
Considering the gap between synthesized and real data, domain adaptation~\cite{li2022annotation,li2021Restoration} was also introduced to further promote the performance of the restoration model based on synthesized data.

Although the previous studies have achieved outstanding performances, the deployment of the restoration algorithms remains a challenging task.
1) It is impractical to collect low-high quality paired cataract images, since the high-quality one is only available from post-operation eyes.
2) Retinal details are always neglected by the model learned from unpaired data, and the deployment of domain adaptation is limited by the requirement of target data.
3) Preserving retinal structures with image guidance, such as segmentation aggravates the requirement of annotations and impacts the robustness of the algorithm.
4) Existing algorithms focus on enhancing image quality, while ignoring the performance improvement of clinical applications from the enhancement.

To address these problems, a cataract restoration network, called SCR-Net, is proposed to enhance cataract fundus images in the absence of supervised data. Specifically, to enforce the training of SCR-Net, SCS is generated by synthesizing cataract fundus images with identical structures according to the imaging principle, and then HFCs are extracted to constrain the structure consistency of SCS as well as enforce the structure preservation in the restoration.  
Our main contributions are summarised as follows:
\begin{itemize}
    \item A restoration network, called SCR-Net\footnote[1]{Code is public available.}, is proposed to enhance the quality of cataract fundus images based on synthesized training data.
    \item According to the fundus imaging principle, a synthesis model of cataract images is proposed, and the cataract images sharing identical structures are hence synthesized to compose the SCS.  
    \item The structure consistency in the HFCs of the SCS is introduced to boost the model training and structure preservation. 
    \item Experiments demonstrate the effectiveness of the proposed approach, by which data requirement is alleviated and superior performance are presented when compared with state-of-the-art algorithms in the cataract image enhancement and the follow-up clinical applications of segmentation and diagnosis.
\end{itemize}

\section{Methodology}
\begin{figure}[t!]
    \begin{centering}
        \includegraphics[width=1\linewidth]{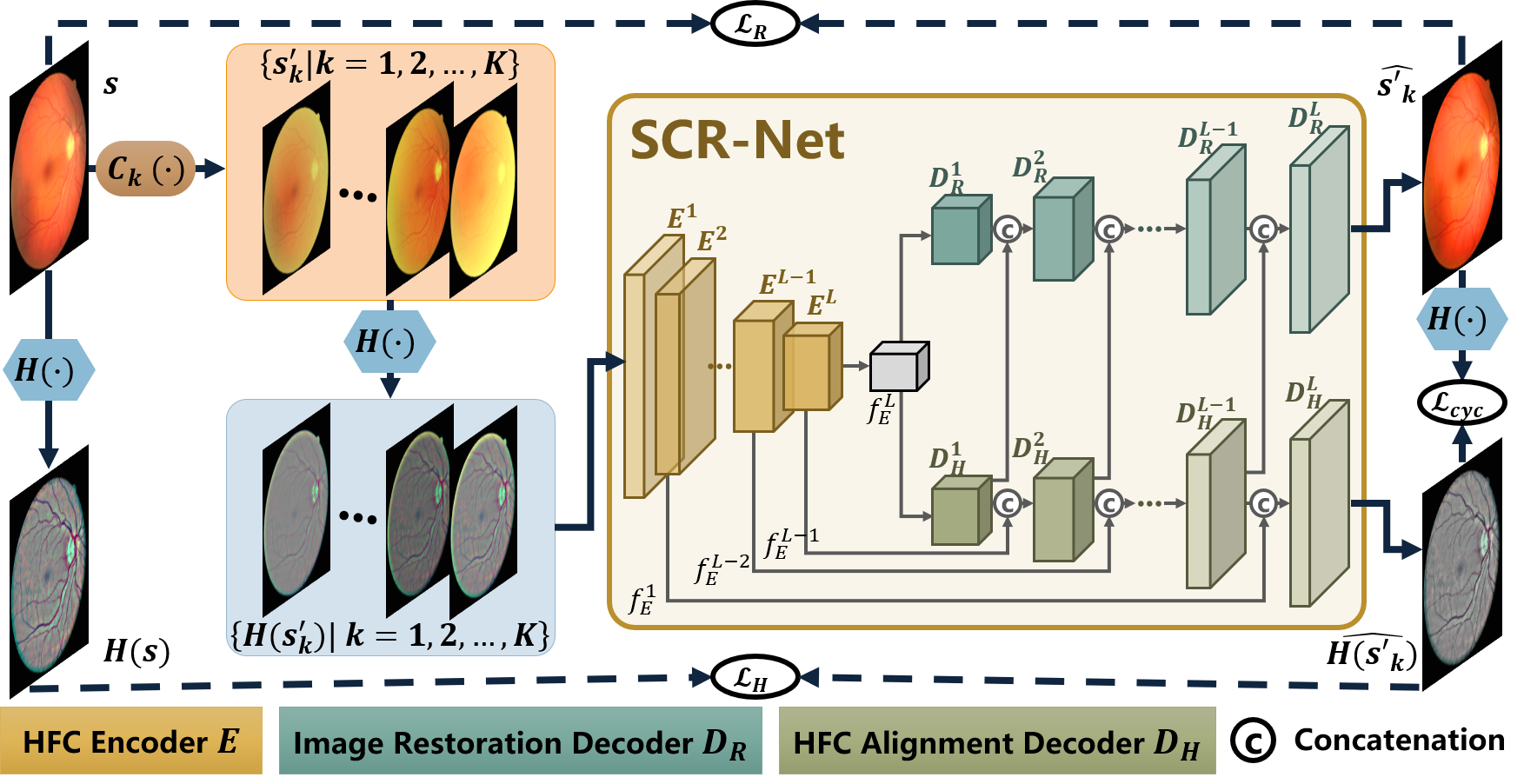}
        \par
    \end{centering}
\caption{Overview of the proposed restoration algorithm. 
From an identical clear image $s$, $C_k(\cdot)$ synthsizes the SCS $\{s'_k\mid k=1,2,...,K\}$ with various cataract parameters, and then $H(\cdot)$ extracts the HFCs from the clear and cataract images for structure alignment as well as image restoration. 
The architecture of SCR-Net is composed of an HFC encoder $E$, an image restoration decoder $D_R$, and an HFC alignment decoder $D_H$. The middle layer features $f^l_E$ of $E$ are forward to $D_H$ for aligning $\widehat{H(s'_k)}$ to $H(s)$. Decoder $D_R$ loads the multi-level features $f^l_H$ from $D_H$ to restore the fundus image.
}
\label{fig:workflow}
\end{figure}

Considering the challenges in enhancing fundus images from cataract patients, the SCR-Net 
is developed to restore cataract images, as shown in Fig.~\ref{fig:workflow}. 
As the foundation of fundus assessment, preserving retinal structures is prioritized in fundus image restoration.
Following this consensus, capturing cataract-invariant features of retinal structures is essential to an efficient restoration model.
Thus an SCS sharing identical structures is firstly acquired by simulating multiple cataract fundus images from an individual clear image.
Subsequently,  through the HFCs in fundus images, structure consistency in the SCS is leveraged to boost the learning of the restoration model and structure preservation. 
Specifically, the proposed SCR-Net imports invariant features from HFC alignment to robustly restore fundus images and preserve retinal structures.

\subsection{Synthesized Cataract Set with Identical Structures} 

In \cite{peli1989restoration}, the fundus imaging through a lens with cataracts is formulated as:
\begin{equation}
I(i,j)=\alpha\cdot L\cdot \gamma(i,j)\cdot t(i,j)+L(1-t(i,j)),
\label{eq:imaging}
\end{equation}
where $I(i,j)$ is the pixel at $(i,j)$ in the fundus image $I$ taken through cataracts, and $\alpha$ denotes the attenuation of retinal illumination caused by cataracts. $L$ is the illumination of the fundus camera, and $\gamma(i,j)$ and $t(i,j)$ are the reflectance function of fundus and the transmission function of cataracts, respectively.

Based on Eq.~\ref{eq:imaging}, we design a cataract simulation model given by:
\begin{equation}
s'_c=\alpha \cdot s_c \ast g_B(r_B,\sigma_B) +\beta \cdot J\ast g_L(r_L,\sigma_L)\cdot (L_c-s_c),
\label{eq:degradation}
\end{equation}
\noindent where $s_c$ and $s'_c$, 
with $c\in \{r,g,b\}$,
are one RGB channel of  
the clear and cataract images, respectively. Notations $g_B$ and $g_L$ represent the Gaussian filters for clear image smooth and cataractous panel, where $g(r,\sigma)$ denotes a filter with a radius of $r$ and spatial constant $\sigma$. Specifically, $r_B, r_L\in\{1,2,3\}$, $\sigma_B, \sigma_L\in[10,30]$.
Parameter $\beta$ is a weight coefficient.
A transmission panel $J_{ij} = \sqrt{(i-a)^{2}+(j-b)^{2}}$ is introduced to model the uneven transmission of $t(i,j)$ with the center of $(a,b)$.
The illumination of a channel $L_c$ is given by the maximum intensity of $s_c$. The simulated cataract image is finally defined as $s'= [s'_r,s'_g,s'_b]$.

According to Eq.~\ref{eq:degradation}, discrepant cataract fundus images can be simulated by changing the parameters, such as $(a,b)$, $\alpha$, and $\beta$. 
As demonstrated in the orange box of Fig.~\ref{fig:Example}, an SCS $\{s'_k=C_k(s)\mid k=1,2,...,K\}$ is thus simulated from a clear fundus image $s$, where $C_k(\cdot)$ denotes the $k$th simulation with random parameters and $K$ was set as 16.
Accordingly, the various cataract images in an SCS share identical structures, which is leveraged to enforce structure preservation in the following sections.

\subsection{High-frequency Components with Structure Consistency} 
\label{sec:models}

\noindent Intuitively, the SCS $\{s'_k\mid k=1,2,...,K\}$ contains the retinal structures consistent with the clear image $s$. 
This structure consistency can be leveraged to boost the restoration model training and structure preservation.

Motivated by the Retinex theory~\cite{li2022annotation}, the blur caused by cataracts is considered as a low-frequency noise. Thus removing the low-frequency components (LFCs) can suppress the fundus image degradation results from cataracts, while the retinal structures in fundus images are carried by the rest HFCs.
Consequently, the structure consistency is imported by the HFCs, which not only suppress the degradation from cataracts, but also preserve the retinal structures.

\begin{figure}[!t]
    \begin{centering}
        \includegraphics[width=0.9\linewidth]{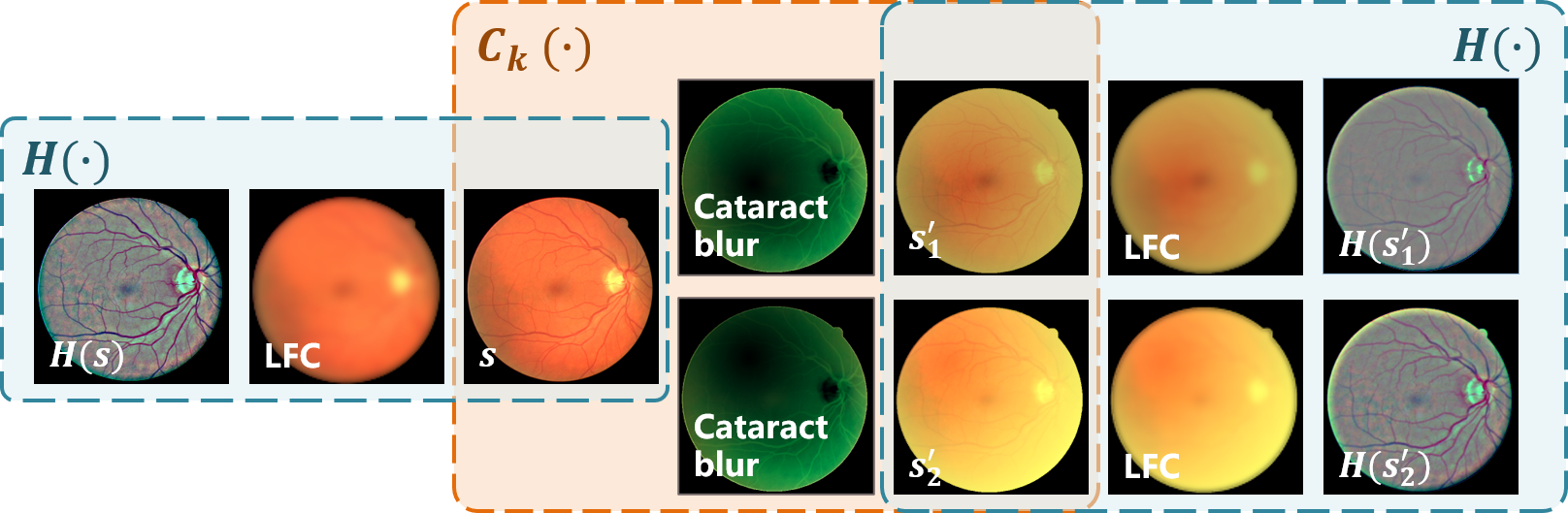}
        \par
    \end{centering}
\caption{Examples for the SCS and HFCs. Randomly simulated cataract blur is imported into the clear image $s$ to acquire the cataract images $s'_1$ and $s'_2$. The HFCs are extracted by removing the LFCs from $s$,  $s'_1$, and $s'_2$.}    
\label{fig:Example}
\end{figure}

The LFCs are extracted by a Gaussian filter, which is a common low-pass filter in signal and image processing. Then the HFCs are straightforwardly captured by removing the LFCs from the fundus image as presented in the turquoise boxes of Fig.~\ref{fig:Example}. This calculation is given by 
\begin{equation}
H(I)= I - I\ast g_P(r_P,\sigma_P),
\label{eq:lowpass}
\end{equation}
\noindent where $H(\cdot )$ denotes the HFCs, and the low-pass Gaussian filter $g_P$ with radius  $r_P$ of 26 and spatial constant $\sigma_P$ of 9, is used to capture the LFCs from image $I$. 
Hence, $\{H(s'_k)\mid k=1,2,...,K\}$ and $H(s)$, denoting the HFCs in the SCS and the clear image, are employed to introduce structure consistency in the training of the restoration network.

\subsection{SCR-Net Architecture} 

\noindent Based on the SCS and HFCs the SCR-Net is hence proposed, in which HFC alignment is implemented to constrain the structure consistency in the HFCs, and features from the alignment are forwarded to optimize the restoration model.
As shown in the khaki box of Fig.~\ref{fig:workflow}, SCR-Net contains three modules. 
The feature encoder $E$, embedding $H(s'_k)$, is shared by the HFC alignment decoder $D_H$ and the image restoration decoder $D_R$. 
Following the structure consistency, decoder $D_H$ aligns $H(s'_k)$ to $H(s)$, and decoder $D_R$ attempts to restore a clear fundus image $\widehat{s'_k}$. 
The latent features from $E$ are forwarded to $D_H$, and the multi-level features in $D_H$ are used in $D_R$.

Decomposing encoder $E$ as a composition of convolution layers, the feature output from a specific layer is given by $f^l_E =E^{l}(E^{l-1}(...E^1(H(s'_k))...)),\;l=1,2,...,L$,
where $l$ denotes the index of layers and $L$ is the total number of layers in $E$.
Decoders $D_H$ and $D_R$ are constructed by the same architecture, which is symmetrical to encoder $E$. The output feature $f^L_E$ of $E$ is forwarded to $D_H$ and $D_R$ as their initial inputs $f^0_H$ and $f^0_R$.
The aligned HFCs is given by $\widehat{H(s'_k)} =D^L_H(f^{L-1}_H)$, where $f^l_H=[D^{l}_H(f^{l-1}_H),f^{L-l}_E]$, $l=1,2,...,L-1$, refers to the concatenation of the outputs of the $l$th later in $D_H$ and the outputs of the symmetrical $(L-l)$th layer in $E$. 
Thus, an alignment loss is calculated to enforce the learning of invariant features, which is given by
\begin{equation}
\mathcal{L}_{H}(E,D_H)=\mathbb{E}\left [ {\textstyle \sum_{k=1}^{K}}\left \| H(s)-\widehat{H(s'_k)} \right \|_{1}\right ].
\label{eq:lh}
\end{equation}
Simultaneously, decoder $D_R$ generates the restored image $\widehat{s'_k}$. 
To construct a robust restoration network, $D_R$ loads $f^l_H$ for reconstructing the restored image with invariant features. Thus $\widehat{s'_k} = D^L_R(f^{L-1}_R)$, where $f^l_R=[D^{l}_R(f^{l-1}_R),D^{l}_H(f^{l-1}_H)]$, $l=1,2,...,L-1$. The restoration loss is given by
\begin{equation}
\mathcal{L}_{R}(E,D_H,D_R)=\mathbb{E}\left[{\textstyle \sum_{k=1}^{K}}\left \| I-\widehat{I'_k} \right \|_{1}\right].
\label{eq:lr}
\end{equation}
Moreover, once the fundus image is properly restored by $D_R$, its HFCs should be consistent with the HFCs aligned by $D_H$. Consequently, a cycle-consistency loss $\mathcal{L}_{cyc}$ is computed between $\widehat{H(s'_k)}$ and the HFCs of $\widehat{s'_k}$ to optimize the network:
%
%
\begin{equation}
\mathcal{L}_{cyc}(E,D_H,D_R)=\mathbb{E} \left[{\textstyle \sum_{k=1}^{K}}\left \| H\left(\widehat{s'_k}\right) - \widehat{H(s'_k)} \right \|_{1}\right].
\label{eq:lcyc}
\end{equation}
The overall objective function is as follows:
\begin{equation}
\mathcal{L}_{total}=\mathcal{L}_{H}(E,D_H)+\mathcal{L}_{R}(E,D_H,D_R)+\mathcal{L}_{cyc}(E,D_H,D_R).
\label{eq:overall}
\end{equation}
Therefore, SCR-Net is optimized with explicit objective function rather than adversarial learning so that the training is easier to achieve convergence. 

\section{Experiments}
\begin{table}[t]
\scriptsize
\centering {\caption{Datasets and evaluation metrics used in the experiments}
\label{tab:dataset} }%
\renewcommand{\arraystretch}{1.2}
\setlength\tabcolsep{4pt}
\begin{tabular}{m{1.4cm}<{\centering}| m{2cm}<{\centering}  m{2cm}<{\centering}|  m{2.5cm}<{\centering}   m{2.7cm}<{\centering} }
\hline
\multirow{2}{1.8cm}{Evaluation} & \multicolumn{2}{m{4cm}<{\centering}|}{With reference} & \multicolumn{2}{c}{Without reference} \\
\cline{2-3} \cline{4-5}
& Restoration & Segmentation & Restoration & Diagnosis \\
\hline
Metrics & SSIM, PSNR & IoU & FIQA & F1-score, Cohen's kappa (Ckappa)\\
\hline
Training set & \multicolumn{2}{m{4cm}<{\centering}|}{DRIVE: 40 clear images} & 300 clear images from Kaggle & 7,331 clear images in Fundus-iSee\\
\hline
Test set & \multicolumn{2}{m{4cm}<{\centering}|}{RCF: 26 pre- and post- operative image pairs} & 100 cataract images from Kaggle & 2,669 cataract images in Fundus-iSee\\
\hline
\end{tabular}
\end{table}

\noindent \textbf{Implementation:} Four fundus image datasets were used to verify the effectiveness of SCR-Net. 
Evaluations and datasets used in the experiment are summarized in Table~\ref{tab:dataset}, where paired training data were generated from the training sets by Eq.~\ref{eq:degradation}. Four kinds of evaluations were implemented.
1) Restoration and 2) segmentation were conducted with the public dataset DRIVE\footnote[2]{http://www.isi.uu.nl/Research/Databases/DRIVE/} and the private dataset RCF to conduct evaluations with reference. The intersection over union (IoU) metric for segmentation was calculated between the restored image and reference.
3) A fundus image dataset from Kaggle\footnote[3]{https://www.kaggle.com/jr2ngb/cataractdataset} was collected to evaluate the restoration without reference, where the fundus image quality assessment (FIQA) from \cite{cheng2021secret} was used as the metric.
4) A diagnostic evaluation was presented with the private dataset Fundus-iSee, which contains 10,000 images sorted into five categories of fundus status. The diagnosis model was learned by ResNet-50 from 5,000 images, and the rest are for testing.


The training data for comparative algorithms were individually synthesized by Eq.~\ref{eq:degradation}. 
The input images had the size $256\times256$, and the training batch size was 8. The model was trained by the Adam optimizer for 150 epochs with a learning rate of 0.001 plus 
50 epochs with the learning rate decaying to 0 gradually. 
The total number of layers in the encoder $E$ was 8. 
The comparison experiment was conducted under the same setting.

\begin{figure}[t!]
    \begin{centering}
        \includegraphics[width=\linewidth]{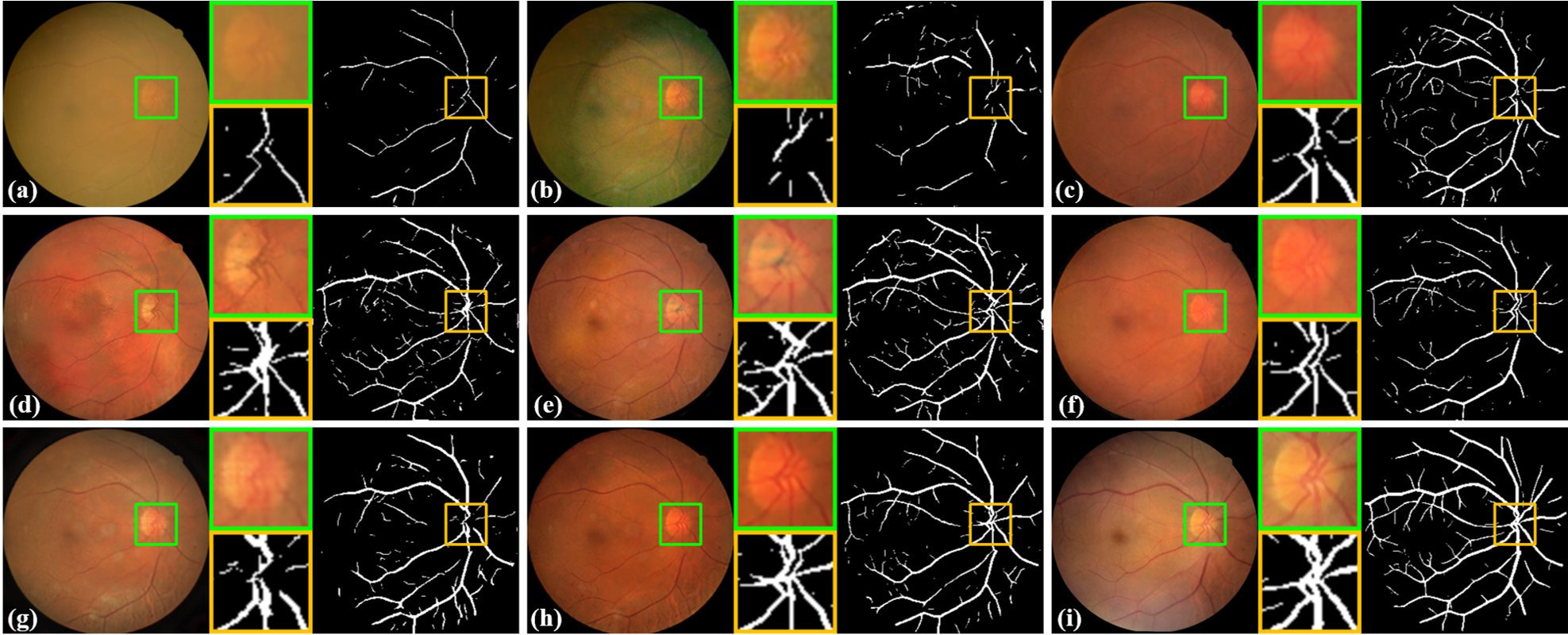}
        \par
    \end{centering}
    \caption{Comparison between the cataract restoration algorithms: (a) cataract fundus image, (b) SGRIF~\cite{cheng2018structure}, (c) CycleGAN~\cite{zhu2017unpaired}, (d) Luo et al.~\cite{luo2020dehaze}, (e) CofeNet~\cite{shen2020modeling}, (f) Li et al.~\cite{li2021Restoration}, (g) I-SECRET~\cite{cheng2021secret}, (h) SCR-Net (ours), and (i) clear image after surgery.} 
    \label{fig:output}
\end{figure}

\noindent \textbf{Comparison and Ablation Study:} To demonstrate the effectiveness of the proposed SCR-Net, comparisons with state-of-the-art methods and an ablation study were conducted. 
The recent methods, SGRIF~\cite{cheng2018structure}, CycleGAN~\cite{zhu2017unpaired}, Luo et al.~\cite{luo2020dehaze}, CofeNet~\cite{shen2020modeling}, Li et al.~\cite{li2021Restoration}, and I-SECRET~\cite{cheng2021secret} were included into comparison.
Table~\ref{tab:dataset} reports the used datasets. 
In the ablation study, the SCS achieved by cataract simulation, the HFCs extracted by $H(\cdot )$, and $D_H$ for domain alignment are respectively removed from the proposed algorithm.
The restoration results are visualized in Fig.~\ref{fig:output} and summarized in Table~\ref{tab:comparison}.

\subsubsection{(1) Comparison}
The comparison of SCR-Net with the competing methods is shown in Fig.~\ref{fig:output}, where the restored images and corresponding segmentation results are exhibited.
Cataracts severely degrade the quality of the example image, and thus only a few of the vessels are identified.
Through image filtering SGRIF~\cite{cheng2018structure} enhances the retinal structures, but the appearance color in the restored image is considerably different from the common ones, which impacts the segmentation. 
The model learned from unpaired data by CycleGAN~\cite{zhu2017unpaired} presents inferior structure preservation.
Guided by segmentation masks, fundus vessels are enhanced by Luo et al.~\cite{luo2020dehaze} and CofeNet~\cite{shen2020modeling}. However, vessel disconnection and pigment disorder are observed in the restored optic disks.
To bridge the gap between the synthesized and real data, domain adaptation is implemented in Li et al.~\cite{li2021Restoration}, which uses both source and target data to train the restoration model. Although domain adaptation boosts the generalization from the source to the target domain,  the access to target data aggravates the burden of data collection.
Though implemented with unpaired data, the importance map in I-SECRET~\cite{cheng2021secret} promises it a decent performance.
The proposed SCR-Net learns a restoration model for cataract images from the synthesized data, and the restoration and segmentation results validate its superior performance on real cataract data.

Quantitative evaluations on restoration, segmentation, and diagnosis are presented in Table~\ref{tab:comparison}. 
As shown in the evaluation metrics, cataracts degrade the fundus image quality and disturb the segmentation and diagnosis.
By applying the restoration algorithms, the quality of cataract images is enhanced and thus the precision of fundus assessment is further boosted.
As a result of the unusual appearance color, though received remarkable results in the diagnosis, SGRIF~\cite{cheng2018structure} receives an inferior assessment in the quantitative metrics of restoration and segmentation.
Mediocre performances are provided by CycleGAN~\cite{zhu2017unpaired} in the segmentation and diagnosis. 
Acceptable performance in the restoration and the follow-up applications is presented by Luo et al.~\cite{luo2020dehaze}.
Among the existing algorithms, by importing structure guidance the recent algorithms, CofeNet~\cite{shen2020modeling}, Li et al.~\cite{li2021Restoration} and I-SECRET~\cite{cheng2021secret} perform prominently, while our SCR-Net has the best performance in all considered evaluation metrics.

\begin{table}[tbp]
\scriptsize
\centering
\caption{Comparisons and ablation study of SCR-Net with state-of-the-art methods on restoration, segmentation, and diagnosis.}
\label{tab:comparison} 
\renewcommand{\arraystretch}{1.15}
\begin{tabular}{p{3.6cm} | | p{1.1cm}<{\centering} p{1.1cm}<{\centering} p{1.1cm}<{\centering} | p{1.1cm}<{\centering} p{1.1cm}<{\centering} p{1.1cm}<{\centering} }
\hline
 &SSIM & PSNR & IoU & FIQA & F1-score & Ckappa \\ 
\hline
Clear & -- & -- & -- & 0.99 & 0.838 & 0.448\\
Cataract & 0.673 & 15.76 & 0.179 & 0.15 & 0.730 & 0.310 \\
\hline
SGRIF~\cite{cheng2018structure} & 0.609 & 15.07 & 0.194 & 0.17 & 0.760 & 0.420\\
CycleGAN~\cite{zhu2017unpaired} & 0.732 & 17.25 & 0.336 & 0.50 & 0.724 & 0.286\\
Luo et al.~\cite{luo2020dehaze} & 0.704 & 17.29 & 0.383 & 0.32 &  0.712 & 0.370\\
CofeNet~\cite{shen2020modeling} & 0.754 & 18.03 & 0.401 & 0.54 &  0.754 & 0.416\\
Li et al.~\cite{li2021Restoration} & 0.755 & 18.07 & 0.376 & 0.52  & 0.747 & 0.405\\
I-SECRET~\cite{cheng2021secret} & 0.748 & 17.63 & 0.380 & 0.46 & 0.734 & 0.382\\
\hline
SCR-Net w/o SCS, $H(\cdot )$, $D_H$ & 0.729 & 17.47 & 0.314 & 0.18 & 0.685 & 0.353\\
SCR-Net w/o $H(\cdot )$, $D_H$ & 0.731 & 17.80 & 0.386 & 0.40 &  0.730 & 0.398\\
SCR-Net w/o $D_H$ & 0.755 & 18.02 & 0.393 & 0.48 & 0.734 & 0.425\\
SCR-Net (ours) & \textbf{0.773} & \textbf{18.39} & \textbf{0.417} & \textbf{0.65}  & \textbf{0.770} & \textbf{0.445}\\
\hline
\end{tabular}
\end{table}

\subsubsection{(2) Ablation Study}
Also from Table~\ref{tab:comparison}, the effectiveness of the proposed three modules of SCR-Net is validated by the ablation study.
The training of the restoration model is enhanced by synthesizing cataract data with sufficient variation. 
Then the HFCs are extracted to preserve retinal structures, as well as constrain structure consistency cooperating with $D_H$. 
Additionally, the objective function of SCR-Net consists of explicit losses so that the model is efficiently optimized. 
Therefore the learned restoration model can be favorably applied to real cataract data.

\section{Conclusion}
Due to the impact of cataracts, diagnosing and monitoring fundus diseases for cataract patients is a challenging task, and supervised data are unavailable to develop restoration algorithms of cataract fundus images.
To increase the certainty of fundus examination, based on structure consistency this paper develops a restoration model for cataract fundus images from synthesized data.
Thanks to its independence from annotations and test data, the proposed algorithm is convenient to deploy in clinics. 
In the experiments, the comparison and ablation study on restoration, segmentation, and diagnosis demonstrate the superior performance and effectiveness of the proposed algorithm.

%
%
\section*{Acknowledgment}
This work was supported in part by The National Natural Science Foundation of China (8210072776), Guangdong Basic and Applied Fundamental Research Fund Committee (2020A1515110286), Guangdong Provincial Department of Education (2020ZDZX3043), Guangdong Provincial Key Laboratory (2020B121201001),  and Shenzhen Natural Science Fund (JCYJ20200109140820699 and the Stable Support Plan Program 20200925174052004).

\bibliographystyle{splncs04}
\bibliography{briefbib}


%




\end{document}